\documentclass{mem}
\usepackage{natbib}\usepackage{txfonts}\usepackage{balance}
\usepackage{graphicx}
\usepackage[a4paper]{hyperref}
\idline{75}{282}
\begin{document}
\def\teff{$T\rm_{eff }$}
\def\kms{$\mathrm {km s}^{-1}$}


\newcommand{\gsim}{\hbox{\rlap{\lower.55ex\hbox{$\sim$}} \kern-.3em
\raise.4ex \hbox{$>$}}}
\newcommand{\lsim}{\hbox{\rlap{\lower.55ex\hbox{$\sim$}} \kern-.3em
\raise.4ex \hbox{$<$}}}

\title{
Tracing differential reddening with Diffuse Interstellar Bands
}
\subtitle{The globular cluster \object{M~4} as a testbed}

\author{
A. \,Monreal-Ibero\inst{1} 
\and
R. Lallement\inst{1} 
\and
L.\, Puspitarini\inst{1,2}
\and
P.\,Bonifacio\inst{1}
\and
L.\,Monaco\inst{3}
\thanks{Based on observations carried out at  the European Southern  Observatory, Paranal (Chile), programmes 077.D-0182, 081.D-0356, and 085.D-0537.}
}
\offprints{A. Monreal-Ibero}

\institute{GEPI, Observatoire de Paris, CNRS UMR811, Universit\'e
   Paris Diderot, Place Jules Janssen, 92190 Meudon, France\\
\email{ana.monreal-ibero@obspm.fr}
\and
Bosscha Observatory and Department of Astronomy, FMIPA, Institut Teknologi Bandung, Jl. Ganesha 10, Bandung 40132, Indonesia 
\and
Departamento de Ciencias Fisicas, Universidad Andres Bello, Rep\'ublica 220, 837-0134 Santiago, Chile
}

\authorrunning{Monreal-Ibero et al.}

\titlerunning{Differential extinction as traced by DIBs in M~4}

\abstract{
Diffuse interstellar bands (DIBs) are weak absorption features of interstellar origin present in the optical and infrared spectra of stars.
Their use as a tool to trace the structure of the Galactic ISM is gaining relevance in the recent years.
Here we present an experiment to test our ability to trace differential reddening on the plane of the sky by using the information relative to the DIB at $\lambda$6614 extracted from the spectra of cool stars. For that we made use of archive FLAMES data of the globular cluster \object{M~4}, as well as \emph{WISE} and \emph{Planck} images for reference.
We found a global positive trend between the distribution of the strength of the DIB, as traced by its equivalent width, and the amount of Galactic reddening, as traced by  \emph{Planck}. This result supports the use of DIBs to trace the small scale structure of the Galactic ISM.
%

}
\maketitle{}

\section{Introduction}

Diffuse Interstellar Bands (DIBs) are non-stellar weak absorption
features found in the spectra of stars that are view through one or several clouds of Interstellar Medium \citep[ISM, see][for a
review]{Sarre06}.
They were identified as unclassified features around the early 20's by \citeauthor{Heger22} and their interstellar origin was established in the 30's
\citep{Merrill34}.
Most DIBs-related studies have been devoted to targets in our Galaxy (see below), although the number of detections of these features in other galaxies of the Local Group \citep[e.g.][]{Welty06,vanLoon13,Cordiner08,Cordiner11} and even further away \citep[][]{Heckman00,MonrealIbero15} is increasing in recent years. 

%
Noteworthy, DIBs present good correlations with the amount of neutral hydrogen along a given line of sight, the extinction and the interstellar Na\,I\,D and Ca\,H\&K lines \citep[e.g.][]{Herbig93,Friedman11}.
Because of this property, these features have recently been started to be used as tracers of the ISM structure \citep[see e.g.][and several contribution in this volume]{Munari08,Yuan12,Kos13,Zasowski15}. This offers certain advantages when used instead of (or in addition to) other methodologies. For example, when comparing with multiband photometry, the use DIB (or any other ISM spectral feature) spectroscopic data provides with additional kinematic information for the ISM. Also, given the intrinsic weakness of the DIBs, these are good features to trace the ISM structure in conditions where other features (e.g. Na\,I\,D) saturate, like very dense molecular clouds or regions seen through a large amount of extinction.

Traditionally, investigations related with DIBs are done using hot (early-type) star spectra. The rationale behind this is clear: hot stars are brighter and present a spectrum dominated by a smooth, feature-less continuum. However, since DIBs seem sensitive to the radiation field of hot stars \citep{Vos11,Dahlstrom13}, these are not \emph{a priori} the optimal targets to prove the typical conditions of the general ISM of our Galaxy. Moreover, hot stars are not automatically abundant (or even existent) in a given region of interest.

Another option, that actually may take the most of the on-going and forthcoming spectroscopic Galactic surveys,  would be using the information carried in cool stars spectra since they are much more numerous and prove less extreme conditions of the ISM in terms of radiation field.
This approximation has been utilized in some recent works. For example, \citet{Yuan12} detected DIBs at $\lambda$5780.4 and $\lambda$6283 in cool-star spectra by reproducing the stellar emission with the help of a set of template spectra made out of a sample of un-reddened stars. A similar approach is utilized by \citet{Kos13} to look for the DIB at $\lambda$8620.4.
Alternatively, the stellar component can be reproduced by means of synthetic stellar models. This approach was used for example by \citet{Zasowski15} for the infrared DIB at 1.53$\mu$m (see contribution in this volume).

With a similar philosophy, \citet{Chen13} developed a method to automatically fit optical spectra to a combination of a stellar synthetic spectra, the atmospheric transmission and a given DIB empirical profile. This method was applied later on by  \citet{Puspitarini15} to study the variation of the DIBs as a function of the distance along the line of sight as well as to study the DIB-extinction relationship in different regions of the Milky Way. 

A complementary aspect is whether DIBs can be used as a tool to trace  small scale spatial variations of the extinction on the plane of the sky (i.e. differential reddening). This is a quite promising use for these features and several strategies have been or are being explored. A first possibility would be the relatively traditional use of hot star spectra \citep[e.g. $\omega$ Centauri,][]{vanLoon09,MaizApellaniz15}. A perhaps more innovative approach is the use of continuous spectrophotometric mapping with integral field units (e.g. Wendt et al., in prep.).
This contribution presents a third approach which is, in turn, a complementary experiment to the one presented by  \citet{Puspitarini15}. Our aim is exploring up to which point we can trace this differential reddening when using the information for the DIBs extracted from the spectra of the much more numerous cool stars.
For that we use a set of FLAMES spectra for the stars in \object{M~4} (\object{NGC~6121}). With a metallicity relatively low
([Fe/H]$=-1.16$, \citet[][]{Harris96}), this is the closest globular cluster to us \citep[$d\sim1.84$~kpc,][]{Braga15}.
The cluster suffers from a large amount of redddening \citep[$E(B-V)=0.37$][]{Hendricks12} and previous studies already suggest a relatively significant amount of differential extinction \citep[][]{Cudworth90,Mucciarelli11}, which makes it a  good candidate for this experiment.

\section{The data}

\subsection{FLAMES spectra}

\begin{table*}
\caption{Summary of utilized spectral data}
\label{thedata}
\begin{center}
\begin{tabular}{lccccccc}
\hline
Prog. Id.  & Instrumental  & Seeing & Air Mass & $S/N$ & N$_\ast$ & Source \\
               & configuration & \\
\hline
077.D-0182 & UVES &  0.8-1.2 & 1.00-1.99   & 100-120 & 166 &\citet{Marino08}      \\ 
081.D-0356 & GIRAFFE/HR15 & $\sim$0.8  & 1.21-1.41 & 50-80 & 86 & \citet{Mucciarelli11} \\ 
085.D-0537 & GIRAFFE/HR15 & 0.9-1.6 & 1.00-1.41 &  70-140 & 99 & \citet{Monaco12} \\ 
\hline
\end{tabular}
\end{center}
\end{table*}

All the optical spectra used in this contribution were taken with the spectrograph FLAMES at the UT2/VLT \citep{Pasquini02} using either the UVES or GIRAFFE configuration and  downloaded from the ESO data archive. The analysis of the stellar component of the spectra has already be published \citep[see ][]{Marino08,Mucciarelli11,Monaco12}  and therefore, the basic stellar parameters (i.e. effective temperature $T_{eff}$, gravity  $\log g$, metallicity [Fe/H], and microturbulence $\xi$) needed to reproduce the stellar emission in the spectral region of the DIB at $\lambda$6614 (see Sec. \ref{secmethod}) are available. Table \ref{thedata} summarizes instrumental configuration, observing conditions, quality and number of used spectra, as well as the reference source for the assumed stellar parameters. 

Individual frames were processed using version 2.13 of the FLAMES/GIRAFFE data reduction pipeline\footnote{http://girbldrs.sourceforge.net/}. The reduction includes bias subtraction, spectra tracing and extraction, correction of fiber and pixel transmission. Additionally, for each frame, sky emission was removed using a high signal-to-noise sky spectrum created by co-adding the spectra of all the sky-dedicated fibers.



\subsection{Planck and WISE reference images}

To test our extraction method, we compared our measurements with other tracers of the 2D structure of the ISM towards \object{M~4}. Specifically, we used archive images from the  \emph{Wide-field Infrared Survey Explorer} (\emph{WISE}) and \emph{Planck} satellites. 

The \emph{WISE} \citep{Wright10} mapped the whole sky in the four infrared bands centered at 3.4, 4.6, 12, and 22~$\mu$m with angular resolutions of 6\farcs1, 6\farcs4, 6\farcs5, and 12\farcs0, respectively. The 3.4 and 12 $\mu$m bands are sensitive to prominent PAH emission features while the 4.6 $\mu$m band traces the continuum emission from very small grains, and the 22 $\mu$m one sees both stochastic emission from small grains and the Wien tail of thermal emission from large grains. We downloaded from the NASA/IPAC Infrared Data Archive\footnote{\texttt{http://irsa.ipac.caltech.edu/}} images corresponding to \object{M~4} as seen with the filter at 12~$\mu$m, the so called W3 band.  A square 4 pointing dither pattern with overlapping areas was used to map the cluster. We estimated the offsets between the images by using several point sources common between each pair of images. Then all the four images were combined with \texttt{imcombine} in IRAF\footnote{The Image Reduction and Analysis Facility IRAF is distributed by the National Optical Astronomy Observatories which is operated by the association of Universities for Research in Astronomy, Inc. under cooperative agreement with the National Science Foundation}.

\emph{Planck} is an ESA mission  operating in the far infrared. Its main aim is mapping the relic radiation from the Big Bang. However, maps at a spatial resolution of $\sim5^\prime$ modeling the all-sky thermal dust emission are an important side-product of the mission \citep{Planck14}.  We downloaded the reddening map around \object{M~4} as predicted by \emph{Planck} (hereafter $E(B-V)_{Planck}$) as well as the corresponding maps for the three modified blackbody parameters ($\tau_{353}$, $T_{obs}$, and $\beta_{obs}$) from the \emph{Planck} Legacy Archive\footnote{\texttt{http://pla.esac.esa.int/pla}}. $E(B-V)_{Planck}$ and $T_{obs}$ will be used in the discussion below.


\section{Fitting method and extraction of DIB parameters \label{secmethod}}

Each observed spectrum was fitted to a model made out of the product of several components as follows:

\begin{eqnarray}
M(\lambda) & =  &S_\lambda[V_{star}] \times T_\lambda[V_{tell}]^{\alpha_{tell}} \times  \\
                    &    &\Pi^i(DIB^i_\lambda[vel^i] \alpha^i) \times ([A]+[B]\times \lambda) \nonumber
\end{eqnarray}

$\bullet$ \emph{$S_\lambda[V_{star}]$,  synthetic stellar spectrum:}
The global shape of this function was derived using the stellar parameters (i.e. $T_{eff}$,  $\log g$,  [Fe/H]) and $\xi$) together with the ATLAS 9 and SYNTHE suite \citep[][]{Kurucz05,Sbordone04,Sbordone05}. Small global variations in wavelength were allowed by introducing the star velocity as a free parameter.

$\bullet$ \emph{$T_\lambda[V_{tell}]^{\alpha_{tell}}$,  synthetic telluric transmission:} 
This term represents the transmission of the earth atmosphere. The global shape of this function was determined by using the facility TAPAS\footnote{\texttt{http://www.pole-ether.fr/tapas/}} \citep{Bertaux14}. Additionally, two free parameters take into account a global scaling factor and the relative movements between Sun and Earth.

$\bullet$ \emph{$\Pi^i(DIB^i_\lambda[vel^i]^{\alpha\,i})$, a product of up to two DIB profiles:}
In this contribution, we worked always with the DIB at $\lambda$6614. The shape of each component was determined empirically as described by \citet{Puspitarini13}. Additionally, for each component, two free parameters characterized its global strength and velocity. 

$\bullet$ \emph{$([A]+[B]\times \lambda)$,  a linear polynomial:}
This term model in a simple fashion the global stellar continuum, requiring two extra free parameters.

The fit was done as a two-steps process. For a given spectrum, we first selected those parts without stellar lines or DIBs and fitted \emph{only} the continuum. Then, we fixed the continuum using the parameters determined in this first step and fitted the other components of the spectrum.  
Further details about the spectrum fitting can be found in \citet{Puspitarini15}. Hereafter, we will use the equivalent width of the DIB (EW($\lambda$6614)) as derived from this fit.


\begin{figure*}[]
\resizebox{\hsize}{!}{\includegraphics[clip=true, angle=270, bb=28 65 555 755]{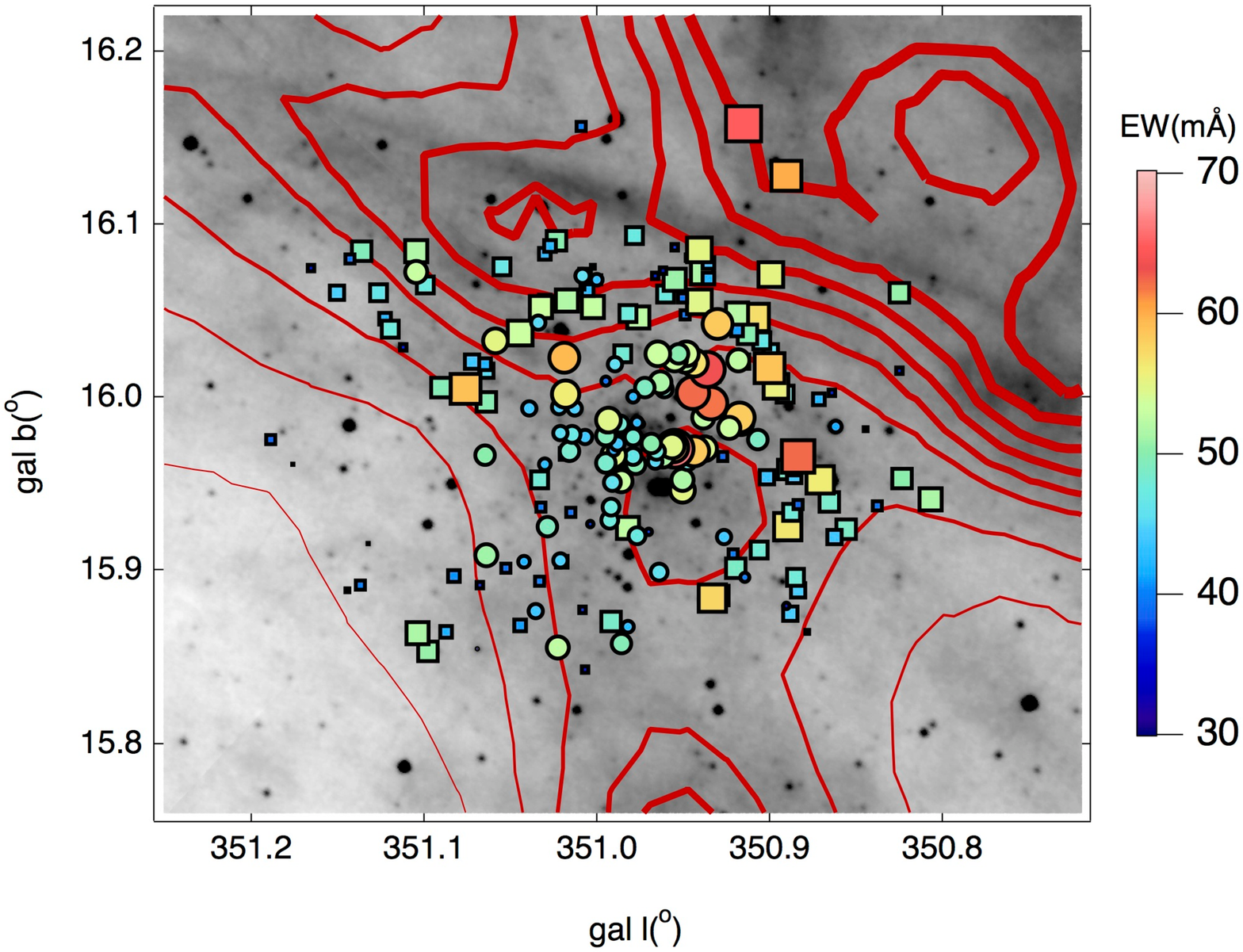}}
\caption{
\footnotesize
Graphic summarizing our results. The positions of the sight of lines (i.e. star spectra) for which we could extract the DIB information are marked with squared (\emph{GIRAFFE}) and circles (\emph{UVES}). Size of the symbols is proportional to the strength of the DIB as traced by its equivalent width. The color bar at the right of the image shows the covered range of equivalent width. Our measurements have been overplotted on top of two images: i) a map in the W3 band of \emph{WISE} is shown as a black and which image; ii) a $E(B-V)$ map as derived from the \emph{Planck} data is shown with contours with a width proportional to the amount of estimated reddening.
}
\label{m4map}%
\end{figure*}

\begin{figure*}[t!]
\resizebox{\hsize}{!}{\includegraphics[clip=true]{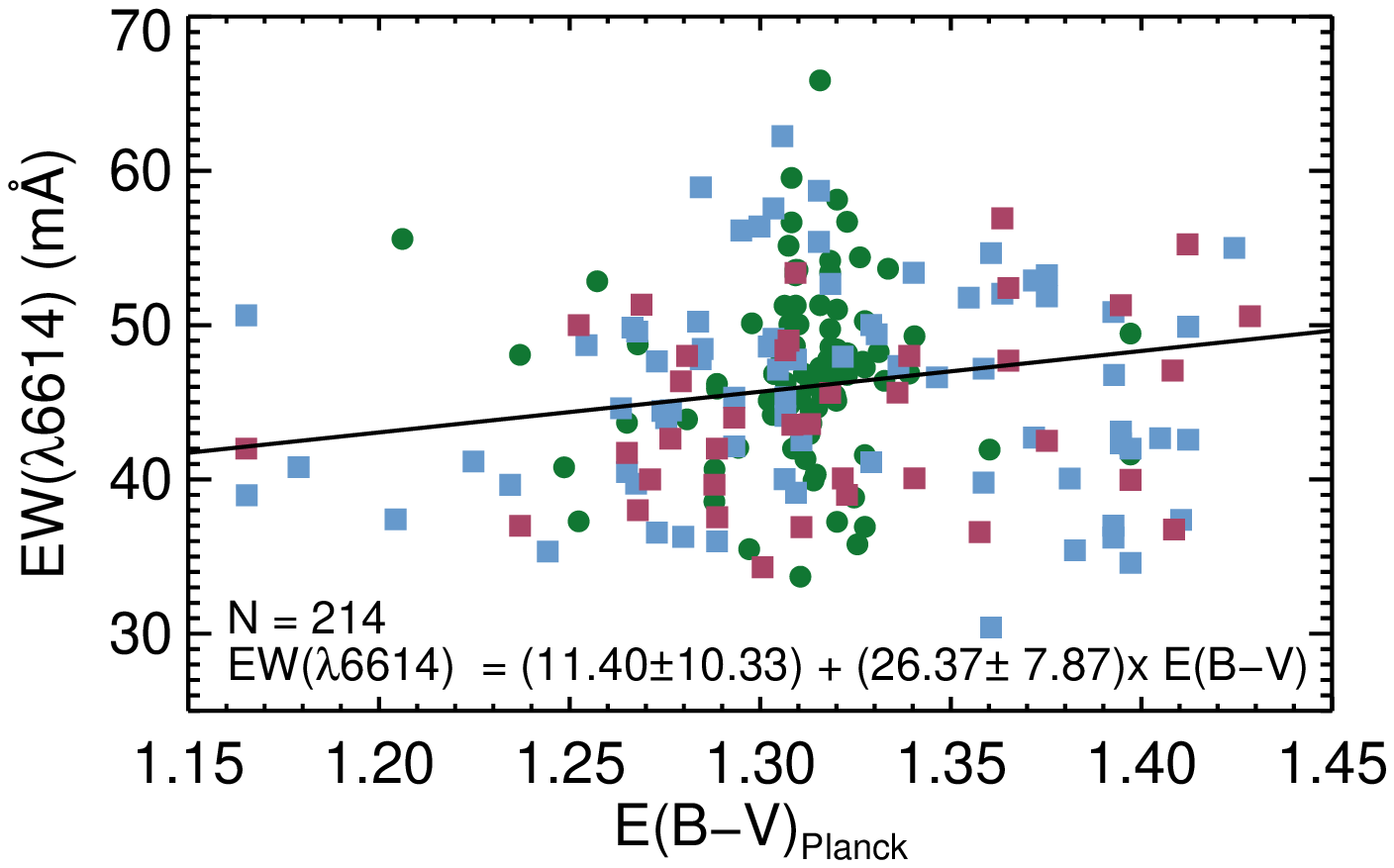}}
\caption{\footnotesize
Relation between the strength of the DIB at $\lambda$6614, as traced by its equivalent width, and the extinction as traced by the reddening derived from \emph{Planck} data. Color/symbol code is as follows: green circles - \citet{Marino08}; burgundy squares - \citet{Mucciarelli11}, blue squares - \citet{Monaco12}.  A linear fit to this relation is shown as a solid black line. Total number of utilized data points and the fitted relation appear at the bottom of the plot.
}
\label{eta}
\end{figure*}

\section{Results}

Our results are summarized in Figs. \ref{m4map} and \ref{eta}. The first figure shows the position of the stars used in our analysis with squares or circles of a size proportional to the EW($\lambda$6614) (see also side bar) on top of two images.
The first one, displayed as a map in greyscale, is the \emph{WISE} W3 band image centered 12~$\mu$m. It traces the dust structure at high resolution. However,  it presents a not negligible contamination due to (stellar) points sources which makes the comparison with our measurements difficult.
Because of that, we also present with red contours the $E(B-V)$ map as derived by \emph{Planck} \citep{Planck14}. This map shows that extinction in the zone associated to M~4 present a complex structure. However, a tendency is seen of global increase of extinction from smaller to larger Galactic longitudes and latitudes (i.e. following a diagonal from bottom left to upper right in Fig. \ref{m4map}). Globally, the strength of the the DIB at $\lambda$6614 follows the same tendency, with a larger number of spectra with high equivalent widths in the upper and rightmost side of the figure. This points towards a relation between these two tracers of the extinction and supports the use of DIBs as an independent tracer of the small scale variations of the extinction.

Fig.~\ref{eta} contains the EW($\lambda$6614) vs. the reddening as predicted by \emph{Planck} for each individual spectrum.  This allows us to explore the tendency in a more quantitative manner.
Note that reddening values predicted by \emph{Planck}   ($E(B-V)_{Planck}\sim1.15-1.45$) are much larger than the reported stellar reddening for \object{M~4}  \citep[$E(B-V)_{star}=0.37$,][]{Hendricks12}. Actually, this is not unexpected given the complexity in estimating the optical reddening from the infrared emission observed by extinction \emph{Planck}.
Specifically, \object{M~4} is located in the direction of the nearby $\rho$ Ophiuchi molecular cloud, an area of active star formation. This implies the existence of local sources of heating photons which cause a complex relation between  
thermal dust emission and the actual stellar extinction \citep[see discussion in Sec. 6.2 of ][]{Planck14}. In particular, for a modified blackbody with a temperature of $\gsim21$~K (as the one estimated towards \object{M~4}, map not shown), \emph{Planck} can overestimate the stellar reddening by a factor $\gsim$2.5, which can explain the difference between $E(B-V)_{Planck}$ and $E(B-V)_{star}$.
Anyway, for the purposes of our experiment, we are interested not in the absolute value but in the relative variations of the $E(B-V)_{Planck}$.
%
Globally there is an increase of EW($\lambda$6614) as $E(B-V)_{Planck}$ increases and the 1-degree polynomial fit to the data is roughly consistent with no having a DIB in conditions of no extinction. However, most of our data points are clustered in the range of $E(B-V)_{Planck}\sim1.25-1.35$, which is relatively small and lacks of enough leverage to establish a firm correlation. Ideally, a search for this would benefit from spectral information in directions with much extreme values of reddening ($E(B-V)_{Planck}<1.25$ or $E(B-V)_{Planck}>1.35$), typically found around the corners of the area displayed in Fig. \ref{m4map}.

\section{Summary and conclusions}

Here we present an experiment to test our ability to trace the differential extinction of the ISM on the plane of the sky by using the information relative to the DIB at $\lambda$6614 extracted from the spectra of cool stars. For that we made use of archive FLAMES data of \object{M~4}, as well as \emph{WISE} and \emph{Planck} images for reference.

Our main conclusion is that we observed a global positive trend between the distribution of the strength of the DIB and the amount of Galactic reddening as traced by  \emph{Planck}. The result is encouraging and supports the use of DIBs to trace the small scale structure of the ISM.

However, the range of DIB strength (and reddening) sampled here is not large enough to offer the necessary leverage to establish a firm correlation between these two parameters. Further support for this result could come from a larger set of spectroscopic observations sampling the more external parts of the cluster, and therefore, areas suffering much extreme amounts of extinction, according to the $E(B-V)_{Planck}$ map.

\begin{acknowledgements}
We acknowledge support from Agence Nationale de la Recherche through the STILISM project (ANR-12-BS05-0016-02).
Based on observations carried out at  the European Southern  Observatory, Paranal (Chile), programmes 077.D-0182, 081.D-0356, and 085.D-0537.
%
This publication makes use of data products from the \emph{WISE}, which is a joint project of the University of California, Los Angeles, and the Jet Propulsion Laboratory/California Institute of Technology, funded by the NASA.
%
Based on observations obtained with Planck (http://www.esa.int/Planck), an ESA science mission with instruments and contributions directly funded by ESA Member States, NASA, and Canada.

\end{acknowledgements}


\bibliography{mybib_aa}{}
\bibliographystyle{./aa}

\end{document}